\documentclass[useAMS,usenatbib]{mnras}  

\usepackage{epsfig}
\usepackage{graphics}
\usepackage{amsmath}
\usepackage{amssymb}
\usepackage{enumerate}
\usepackage{natbib}
\usepackage{latexsym}
\title{Primordial black hole evolution in two-fluid cosmology}

\author[E. M. Guti\'errez, F. L. Vieyro, and G. E. Romero]{E. M. Guti\'errez$^{1}$\thanks{E-mail: emgutierrez@iar.unlp.edu.ar}, F. L. Vieyro$^{1,2}$, and G. E. Romero$^{1,3}$\\
$^{1}$Instituto Argentino de Radioastronom\'{\i}a (IAR, CCT La Plata, CONICET; CICPBA), C.C.5, (1984) Villa Elisa, Buenos Aires, Argentina\\
$^{2}$Departament de F\'{i}sica Qu\`antica i Astrof\'{i}sica, Institut de Ci\`encies del Cosmos (ICCUB), Universitat de Barcelona, IEEC-UB, \\Mart\'{i} i Franqu\`es 1, E08028 Barcelona, Spain \\
$^{3}$Facultad de Ciencias Astron\'omicas y Geof\'{\i}sicas, Universidad Nacional de La Plata, Paseo del Bosque s/n, 1900, La Plata, Argentina
}

\begin{document}

\date{Accepted 2017 October 06. Received 2017 October 06; in original from 2017 August 29} 

\pagerange{L\pageref{firstpage}--L\pageref{lastpage}}

\maketitle

\label{firstpage}

\begin{abstract}

Several processes in the early universe might lead to the formation of primordial black holes with different masses.
These black holes would interact with the cosmic plasma through accretion and emission processes. Such interactions might have affected the dynamics of the universe and generated a considerable amount of entropy.
In this paper we investigate the effects of the presence of primordial black holes on the evolution of the early universe. We adopt a two-fluid cosmological model with radiation and a primordial black hole gas.
The latter is modelled with different initial mass functions taking into account the available constraints over the initial primordial black hole abundances.
We find that certain populations with narrow initial mass functions are capable to produce significant changes in the scale factor and the entropy.

\end{abstract}


\section{Introduction}

	There is overwhelming evidence supporting the existence of black holes in the universe. Studies of stellar and gas dynamics strongly suggest the presence of supermassive black holes ($M_{\rm BH} \sim 10^6-10^9 ~ \rm M_{\odot}$) at the centre of most galaxies \citep[e.g.][]{ferrarese2005}. At smaller scales, stellar-mass black holes ($M_{\rm BH} \sim 3-10 ~ \rm M_\odot$) are thought to be the result of the  collapse of massive stars \citep{neugebauer2003}. Their presence is manifested in X-ray binaries (XRBs). There are currently about  60 stellar black hole candidates \citep{corral2016}.
Additionally, recent gravitational wave detections have revealed the existence of binary systems of black holes with several tens of solar masses at moderate redshifts \citep{abbott2016a, abbott2016b, abbott2017}.

	Under the extreme conditions of the early universe, black holes can be formed from direct collapse \citep{zeldovich1966,hawking1971,carr1974}. These are called \textit{Primordial Black Holes} (PBHs) and have been extensively investigated (see \citealt{khlopov2010} for a review). 
	
	A PBH would form with a mass of order the horizon mass $M_{\rm H}(t)$,
\begin{equation}
	M_{\rm PBH} \sim M_{\rm H}(t) \sim \frac{c^3 t}{G} \sim 10^{15} \left( \frac{t}{10^{-23} ~ \mathrm{s}} \right) ~ \mathrm{g},
\end{equation}
as can be seen by a simple comparison between the cosmic density at time $t$ after the Big Bang and the density associated with a black hole of mass $M$.
This implies that PBHs could span a wide range of masses. In particular, they might be small enough for Hawking radiation to be important \citep{hawking1974}.

  PBH formation requires the existence of large inhomogeneities in the early universe \citep{carr1974, carr1975}. Independently of the source of these inhomogeneities, the formation can be enhanced by some processes, such as phase transitions --for example from bubble collisions \citep{crawford1982, hawking1982}, collapse of cosmic strings \citep{hawking1987, polnarev1991} or domain walls \citep[e.g][]{berezin1983, caldwell1996}-- or a sudden reduction in the pressure at the quark-hadron era \citep{jedamzik1997, jedamzik1999}.
Furthermore, applications of ``critical phenomena'' to PBH formation suggest that their spectrum could go well below the horizon mass \citep[e.g.][]{niemeyer1998, green1999}. PBH formation can also occur in a matter-dominated universe \citep[see, e.g.,][]{khlopov1980, polnarev1985}. For more details on all these mechanisms the reader is referred to \citet{carr2010}.

	Although PBHs have not been detected so far, their study covers several areas of interest: baryogenesis \citep{barrow1980, lindley1981, barrow1991b, hook2014}, dark matter (see \citealt{carr2016a} and references therein), Big Bang nucleosynthesis \citep{zeldovich1977, vainer1978, kohri2000}, reonization of the universe \citep{gibilisco1998}, and gravitational waves \citep{bird2016, sasaki2016}. 
	
	Several constraints have been imposed on the initial number of PBHs formed (see \citealt{carr2010} for a review). Most of these limits are related to the different potential interactions between PBHs and other astrophysical objects (e.g. gravitational interactions) and the observables resulting from black hole evaporation. The importance of these constraints is that they indirectly impose restrictions on the conditions of the early universe and, hence, on different early universe models (e.g. inflation models, \citealt{josan2010, peiris2008}). However, not all constraints on PBHs  are equally reliable. For example, at the lowest masses ($M \lesssim 10^6 ~ \mathrm{g}$) the only available constraint relies on a strong assumption, namely that black holes do not completely evaporate but leave behind Planck-mass relic particles \citep{macgibbon1987}.

	The presence of a PBH population in the early universe could have affected the cosmic evolution directly. The main feature of such a population is its \textit{Initial Mass Function} (IMF). According to the particular mechanism and timescale of the formation process, this IMF can either extend over a wide mass range or be narrow and centred on a certain mass. \citet{barrow1991} studied the cosmic evolution of the early universe considering radiation and a population of PBHs  with a power-law IMF; they assume that the two components interacted only through Hawking evaporation. Other scenarios considered in later works involve PBH populations with narrow or {\it monoenergetic} IMFs \citep[e.g.][]{barrow1992,zimdahl1998,brevik2003} or additional space-time dimensions \citep{borunda2010}.

  In this work, we investigate the early universe evolution considering PBH populations with both extended and narrow IMFs, taking into account the best available constraints on PBH abundances in the characterization of the scenarios. We consider a FLRW space-time and a two-perfect-fluid model: a PBH gas with a dust-like equation of state, and a relativistic component (radiation). The fluids exchange energy through Hawking evaporation and accretion on to the black holes. The energy exchange is coupled to the metric scale factor through the Friedmann equations.

	The structure of this article is the following: in Sec. \ref{Sec:abundances} we briefly summarise the most significant available constraints about PBH abundances. In Sec. \ref{Sec:interactions} we analyse the energy exchange between a black hole and a radiation bath. Then, in Sec. \ref{Sec:evolution} we extend this analysis to a black hole population interacting with radiation in a cosmological background. We present the results in Sec. \ref{Sec:results}, and the conclusions and final remarks in Sec. \ref{Sec:conclusions}.

\section{PBH abundances}
\label{Sec:abundances}

	The constraints on the initial abundances of PBHs are generally expressed in terms of the fraction of the energy density of the universe that goes to PBHs at their formation epoch: $\beta = \rho_{\rm PBH} / \rho_{\rm tot}$.

The lifetime of a black hole with mass $M$ due to Hawking evaporation can be estimated as \citep{hawking1975}
\begin{equation}
  \tau_{\rm life} \sim 10^{10} \Big( \frac{M}{10^{15} \textrm{ g}} \Big) \textrm{ yr}.
\end{equation}
	As described in \citet{carr2010}, this implies that \textit{a)} PBHs formed with mass of order of $10^{15} ~ \rm g$ should be evaporating at the present epoch producing gamma rays, positrons, and antiprotons that contribute to the diffuse gamma ray background and the cosmic ray flux \citep[e.g.][]{wright1996, carr2016b}.
\textit{b)} PBHs with initial mass $M < 10^{15} ~ \mathrm{g}$ are already evaporated; however, their existence could have affected different processes in the early universe. Ones that evaporated within the first second after the Big Bang could have generated most of the entropy of the universe \citep[e.g.][]{zeldovich1976} or altered the baryogenesis \citep{dolgov2000,bugaev2003} and the Big Bang nucleosynthesis \citep[e.g.][]{zeldovich1977,vainer1978}.
These PBHs can also evaporate into neutrinos, hadrons, and other massive particles, or leave behind Planck-mass relic particles contributing to the cold dark matter (CDM) \citep[e.g.][]{bugaev2002, lemoine2000, macgibbon1987, barrow1992, alexander2007}.
\textit{c)} PBHs with $M > 10^{15} ~ \mathrm{g}$ have lifetimes longer than the age of the universe, and hence they would still exist and would be detectable by their gravitational effects.
Indeed, given the negative results obtained so far in the search for particle dark matter  (in particular weakly-interacting massive particles, \citealt{akerib2016}), PBHs have become interesting CDM candidates \citep[see, e.g.,][]{chapline1975,carr2016a}.
In addition, these PBHs can interact with other astrophysical objects in several ways; for example, they might be captured by a neutron star and the star being accreted \citep{capela2013}, or they could have played a role as seeds of supermassive black holes in the centre of galaxies. Recently, they have been also proposed as sources of gravitational wave events \citep{garcia2017}.

\begin{figure}
	\centering
	\includegraphics[width=0.45\textwidth,keepaspectratio]{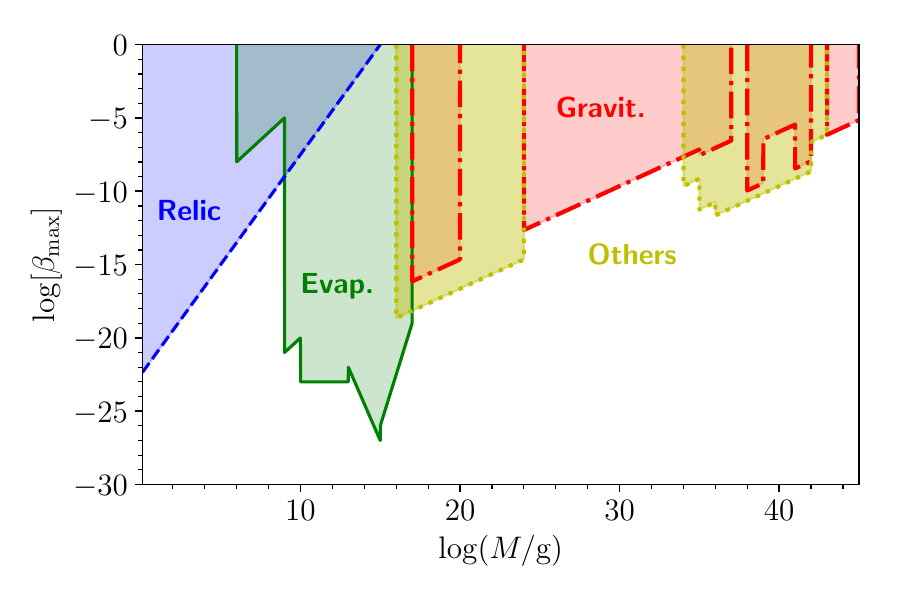}
\caption{ Most relevant constraints on the initial fraction of the energy density of the universe in PBHs for monoenergetic IMFs (adapted from \citealt{green2014}). We distinguish in colours the different kind of constraints. The constraints derived from relic particle studies are shown in blue, those related to evaporation effects (e.g. gamma ray production) in green, those related to gravitational effects (e.g. lensing or dynamical effects) in red, and other constraints, as those related to neutron capture or gravitational waves, in yellow. The coloured areas are the forbidden regions of the parameter $\beta=\rho_\mathrm{PBH}/\rho_\mathrm{tot}$.}
\label{fig:constraints}
\end{figure}

	Since these effects are not observed, constraints on the PBH abundances are imposed in accordance with the sensitivity of current observations. Figure \ref{fig:constraints} shows the most updated limits. The constraints on black hole relics are the only ones that can limit the quantity of less massive PBHs formed.
It is important to remark, as was done by \citet{carr2017} and \citet{kuhnel2017}, that the constraints in Fig. \ref{fig:constraints} are derived considering monoenergetic PBH populations. The case of PBHs with an extended mass function is quite different and depends on the particular form of the IMF. Following \citet{carr2017}, we summarise the main aspects of this treatment in Appendix \ref{ap:constraints_extended}.

\section{Black hole mass variation}\label{Sec:interactions}

	A Schwarzschild black hole of mass $M$ emits particles with a blackbody spectrum at a temperature \citep{hawking1975}
	
\begin{equation}
	T_{\rm BH}= \frac{\hbar c^3}{8 \pi k_{\rm B}GM},
\end{equation}

\noindent where $\hbar$ is the reduced Planck constant, $k_{\rm B}$ is the Boltzmann constant, and $G$ is the gravitational constant. This emission produces a mass loss rate given by

\begin{equation}
\label{eq:masa_1}
	\frac{\partial M}{\partial t} \bigg \vert_{\rm em} = \frac{1}{c^2}\Sigma_{\rm BH} h_*\sigma T_{\rm BH}^4,
\end{equation}

\noindent where $\sigma$ is the Stefan-Boltzmann constant, $c$ is the speed of light in vacuum, $h_*$ is the number of particle species available for the black hole to evaporate into, and $\Sigma_{\rm BH}$ is the effective area at which particles escape from the hole.
Strictly speaking, $h_*$ depends on the temperature and, consequently, on the black hole mass; nevertheless, this dependency is discrete and $h_*$ can be considered constant for fixed mass ranges. Equation (\ref{eq:masa_1}) can be written as

\begin{equation}
	\frac{\partial M}{\partial t} \bigg \vert_{\rm em} = - \frac{A(M)}{M^2},
\end{equation}

\noindent where $A(M)=5.3 \times 10^{25} \ \mathrm{g}^3 \ \mathrm{s}^{-1}$ for black holes with masses $> 10^{17} \ \mathrm{g}$ and $A(M) \geq 7.8 \times 10^{26} \ \mathrm{g}^3 \ \mathrm{s}^{-1}$ for black holes with masses $\leq 10^{15} \ \mathrm{g}$ \citep{macgibbon1991}.

In addition to the Hawking emission, black holes immersed in a thermal bath accrete particles at a rate given by \citep{zeldovich1966}

\begin{equation}
	\frac{\partial M}{\partial t} \bigg \vert_{\rm acc} = \frac{27\pi G^2}{c^5} \rho_{\rm R} M^2,
\end{equation}

\noindent where $\rho_{\rm R}$ is the energy density of radiation measured far away from the hole. Combining the two effects we obtain the complete equation for the mass variation rate:

\begin{equation}
	\label{eq:tasa_masa}
	\frac{dM}{dt} = \frac{27\pi G^2}{c^5} \rho_{\rm R} M^2 - \frac{1}{c^2} \Sigma_{\rm BH} h_*\sigma T_{\rm BH}^4.
\end{equation}

\section{Cosmic evolution}\label{Sec:evolution}

\subsection{Two-fluid cosmology}

	We consider a cosmic fluid immersed in a FLRW space-time, ($\Sigma \times \mathbb{R},g_{\mu \nu}^\mathrm{FLRW})$, where $\Sigma$ is a set of space-like hypersurfaces and $g_{\mu \nu}^\mathrm{FLRW}$ is such that
\begin{equation}
\label{eq:FLRW_metric}
	ds^2 = -dt^2 + R(t)^2 \left[ \frac{dr^2}{1-kr^2} +r^2 (d\theta^2 + \sin^2 \theta d\phi^2) \right].
\end{equation}	
Here, $R(t)$ is the scale factor of the metric and $k$ is the curvature parameter. We assume that the fluid is composed of two perfect components, $A$ and $B$; hence, its energy-momentum tensor is
\begin{equation}
	T^{\mu \nu} = T_{(A)}^{\mu \nu} + T_{(B)}^{\mu \nu},
\end{equation}	
where
\begin{equation}
	T^{\mu \nu}_{(i)} = \frac{1}{c^2}\left[\rho_{(i)}+p_{(i)}\right] u^\mu u^\nu - p_{(i)} g^{\mu \nu},\quad (i=A,B),
\end{equation}
and $\rho_{(i)}$ and $p_{(i)}$ denote the energy density and the pressure of the fluid-component $i$, respectively. We assume that both components have the same four-velocity $u^{\mu}$ but their equation of state can be different:
\begin{equation}
	p_{(i)} = w_{(i)} \rho_{(i)}, \quad \quad (i=A,B).
\end{equation}

	If the two fluid-components interact, only the total energy-momentum tensor is conserved. Thus,
\begin{equation}
\label{eq:cons_tmunu_tot}
	\nabla_\mu T^{\mu \nu} = 0 \quad \Longrightarrow \quad \nabla_\mu T^{\mu \nu}_{(A)} = - \nabla_\mu T^{\mu \nu}_{(B)}.
\end{equation}
We denote $Q$ the rate of energy exchange caused by the interaction, and we define the normalised scale factor $a(t) = R(t) / R(t_0)$, where $t_0$ is an arbitrary cosmic time (see Sec. \ref{Sec:results} for the specific choice of this value). Then, by adding one of the Friedmann equations to the only non-trivial component of Eq. (\ref{eq:cons_tmunu_tot}) we obtain a system of three differential equations for $\rho_{(A)}$, $\rho_{(B)}$, and $a$:

\begin{equation}
\label{eq:ecuaciones}
\begin{aligned}
	\dot{\rho}_{(A)} + 3\left( \frac{\dot{a}}{a} \right) \left[ 1+w_{(A)} \right] \rho_{(A)} = Q, \\
	\dot{\rho}_{(B)} + 3\left( \frac{\dot{a}}{a} \right) \left[ 1+w_{(B)} \right] \rho_{(B)} = -Q, \\
	\left( \frac{\dot{a}}{a} \right)^2 - \frac{8 \pi G}{3c^2} \left[ \rho_{(A)}+\rho_{(B)} \right] + \frac{c^2 k}{[R(t_0)a]^2} = 0.
\end{aligned}
\end{equation}

\noindent The interaction term $Q$ depends on the specific characteristics of the system. In what follows, we apply this two-fluid formalism to a cosmological model of the early universe in which the two fluids are radiation and a PBH gas.

\subsection{Early universe with PBHs}

Let us consider a relativistic thermal plasma characterised by its equilibrium temperature $T_{\rm R}$, which sets the other relevant thermodynamic quantities (energy density $\rho_{\rm R}$, pressure $p_{\rm R}$, and entropy density $s_{\rm R}$) through the following relations:
\begin{equation}
\label{eq:tq_rad}
\begin{aligned}
	\rho_{\rm R} = g_* \frac{(k_{\rm B}T_{\rm R})^4}{(\hbar c)^3}, \quad \quad \quad \quad \\
	p_{\rm R} = \frac{1}{3}\rho_{\rm R} = g_* \frac{(k_{\rm B}T_{\rm R})^4}{3(\hbar c)^3}, \quad \\
	s_{\rm R} = \frac{\rho_{\rm R} + p_{\rm R}}{T_{\rm R}} = \frac{4}{3}\frac{g_*k_{\rm B}^4}{(\hbar c)^3} T_{\rm R}^3.
\end{aligned}
\end{equation}

\noindent Here $g_*$ takes into account the contribution of the different species of relativistic particles.

	Let us also consider a PBH component modelled as a dust-like perfect fluid ($p_{\rm PBH}=0$) whose constituents are Schwarzschild black holes.
These may have different masses and therefore are characterised by their IMF, $N_{\rm 0}(m)$, which evolves with time due to two processes: the expansion of the universe and the energy exchange of each PBH with the radiation. If $N(t;m)$ denotes the mass function at time $t$, the PBH energy density is

\begin{equation}
	\rho_{\rm PBH}(t) = \int_{M_{\rm min}}^{M_{\rm max}} N(t;m) E(m) dm,
\end{equation} 

\noindent where $M_{\rm min}$ and $M_{\rm max}$ are the minimum and maximum mass of the black holes, and $E(m)=mc^2$ is the energy of a Schwarzschild black hole of mass $m$. In a similar way, the entropy density can be calculated as

\begin{equation}
	s_{\rm PBH}(t) = \int_{M_{\rm min}}^{M_{\rm max}} N(t;m) S(m) dm,
\end{equation}

\noindent where $S(m) = 4\pi k_{\rm B} G m^2 / \hbar c $ is the Bekenstein-Hawking entropy of a Schwarzschild black hole of mass $m$.

In order to obtain an expression for the $Q$-term, we must sum the effects of the interaction of each black hole with the radiation.
If the mass of a black hole evolves from $m$ at time $t$ to $m+dm$ at time $t+dt$, then
\begin{equation}
	N(t;m) = N(t+dt;m+dm),
\end{equation}
and this implies
\begin{equation}
	\frac{\partial N(t;m)}{\partial t} \bigg \vert_{\rm int} = \frac{\partial N(t;m)}{\partial m} \frac{dm}{dt},
\end{equation}
where the mass variation rate is given by Eq. (\ref{eq:tasa_masa}).
The $Q$-term results
\begin{equation}
\begin{aligned}
	Q = \frac{\partial \rho_{\rm PBH}}{\partial t} \bigg \vert_{\rm int} = \int_{M_{\rm min}}^{M_{\rm max}} \frac{\partial N(t;m)}{\partial t} \bigg \vert_{\rm int} mc^2 dm \\
	= \int_{M_{\rm min}}^{M_{\rm max}} \frac{\partial N(t;m)}{\partial m} \frac{dm}{dt} mc^2 dm.
\end{aligned}
\end{equation}
Finally, we consider that the space-time has negligible curvature (which is a very reasonable assumption in the early universe) and we set $k=0$.
The system of equations (\ref{eq:ecuaciones}) becomes
\begin{equation}
\label{eq:system}
\begin{aligned}
	\dot{\rho}_{\rm R} + 4\frac{\dot{a}}{a} \rho_{\rm R} = -\int_{M_{\rm min}}^{M_{\rm max}} \frac{\partial N(t;m)}{\partial m} \frac{dm}{dt} mc^2 dm, \\
	\dot{\rho}_{\rm PBH} + 3 \frac{\dot{a}}{a} \rho_{\rm PBH} = \int_{M_{\rm min}}^{M_{\rm max}} \frac{\partial N(t;m)}{\partial m} \frac{dm}{dt} mc^2 dm, \quad \\
	\left( \frac{\dot{a}}{a} \right)^2 = \frac{8 \pi G}{3c^2} \left( \rho_{\rm R}+\rho_{\rm PBH} \right). \quad \quad \quad \quad \quad \quad \quad \ \
\end{aligned}
\end{equation}
Now we have an integro-differential equation system for the functions $N(t;m)$, $\rho_{\rm R}(t)$, and $a(t)$.
In what follows, we separate our analysis into narrow IMFs and extended IMFs.

\subsubsection{Narrow IMF}

	If a PBH population is formed on a short timescale, for example from a phase-transition, the IMF is typically narrow and centred on a particular mass \citep{barrow1992}.
We study the simplest  narrow IMF, namely a Dirac delta function. We assume that the PBHs form at time $t_{\rm form}$ with a mass $M_{\rm form} \sim M_{\rm H}(t_{\rm form})$. Thus,
\begin{equation}
	N_0 (m) \equiv A \delta (m-M_{\rm form}),
	\label{eq:delta}
\end{equation}

\noindent where $A$ is a normalisation constant and can be related to the original fraction of the energy density of the universe that goes to PBHs.
The initial energy and entropy densities of the black holes are
\begin{eqnarray}
	\rho_{\rm PBH}(t=t_{\rm form}) &= A M_{\rm form}c^2, \\
	s_{\rm PBH}(t=t_{\rm form}) &= A S(M_{\rm form}).
\end{eqnarray}
In this scenario, all black holes evolve in the same manner and at each time $t$ they have the same mass $M_{\rm PBH}(t)$.
Therefore, we can determine the evolution of the whole population by studying one representative PBH. 
Under this simplification, the set of Eqs. (\ref{eq:system}) becomes
\begin{equation}
	\begin{aligned}
		\dot{\rho_{\rm R}}+4 \frac{\dot{a}}{a}\rho_{\rm R} = -Ac^2 \frac{dM_{\rm PBH}}{dt}, \\
		\dot{\rho_{\rm PBH}}+3\frac{\dot{a}}{a}\rho_{\rm PBH} = A c^2 \frac{dM_{\rm PBH}}{dt}, \\
		\left( \frac{\dot{a}}{a} \right) = \frac{8 \pi G}{3c^2} \left( \rho_{\rm R} + \rho_{\rm PBH} \right),
	\end{aligned}
\end{equation}
that is a system of linear differential equations for $M_{\rm PBH}(t)$, $\rho_{\rm R}(t)$, and $a(t)$.

\subsubsection{Extended IMF}

In other scenarios, for example those where the formation occurs on a long timescale, the IMF can be extended and span a wide mass range \citep{barrow1991}. We analyse the particular case of a power-law IMF of the form
\begin{equation}
	N_0(m) = A m^{-\gamma},
	\label{eq:powerlaw}
\end{equation}
where $A$ is a normalisation constant and $\gamma$ is the spectral index, which typically lies in the range $2-3$ \citep{carr2010}.

Here, we assume that the formation starts at time $t_{\rm ini}$ (the least massive black holes) and ends at time $t_{\rm end}$ (the most massive ones).
The form of the mass function $N(t;m)$ varies with time owing to the different rates of evolution of each PBH.
In order to solve the equation system, we discretise the function $N(t;m)$ in blocks.
Each block evolves as an independent Dirac delta function like the one we previously considered.

\section{Results}\label{Sec:results}

	We set the initial values of the energy densities of both fluids at time $t_{\rm ini}$, and we normalise the scale factor such that $a(t_0=t_{\rm ini})=1$.
For the initial radiation temperature, we assume the following expression (e.g. \citealt{weinberg1972}):
\begin{equation}
	T_{\rm R}(t_{\rm ini}) \sim 10^{10} (t_{\rm ini}/\mathrm{s})^{-1/2} \ \mathrm{K}.
\end{equation}
Then, we set
\begin{equation}
	\rho_{\rm PBH}(t_{\rm ini}) := \beta \rho_{\rm R}(t_{\rm ini}),
\end{equation}
where $\beta < 1$ is a free parameter of the model, though it is limited by the constraints previously discussed.
As the number density of pre-inflation PBHs is negligible after the inflationary epoch, we only consider post-inflationary times: $t_{\rm ini}=10^{-33} \ \mathrm{s}$.

	In what follows, we present numerical results for some examples of the two scenarios discussed in Sec. \ref{Sec:evolution}. We solved the equation system (\ref{eq:system}) using an adapted fourth-order Runge-Kutta method.
	
\subsection{Monoenergetic IMF}

\begin{figure}
	\centering
	\includegraphics[width=0.45\textwidth,keepaspectratio]{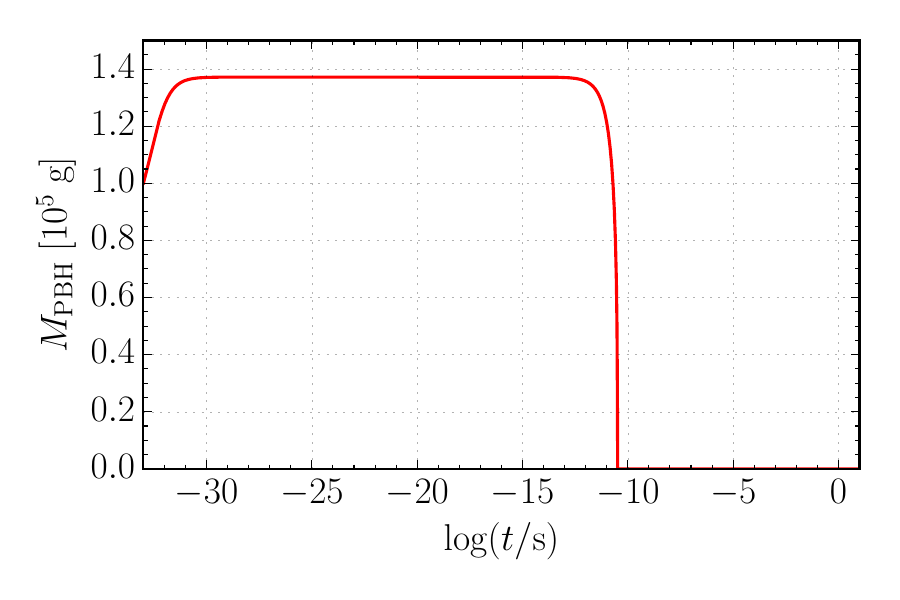}
\caption{Evolution of a PBH of mass $10^5 \ \mathrm{g}$ formed at a time $t_{\rm ini}=10^{-33} \ \mathrm{s}$}
\label{fig:BH_mass}
\end{figure}

\begin{figure}
	\centering
	\includegraphics[width=0.4\textwidth,keepaspectratio]{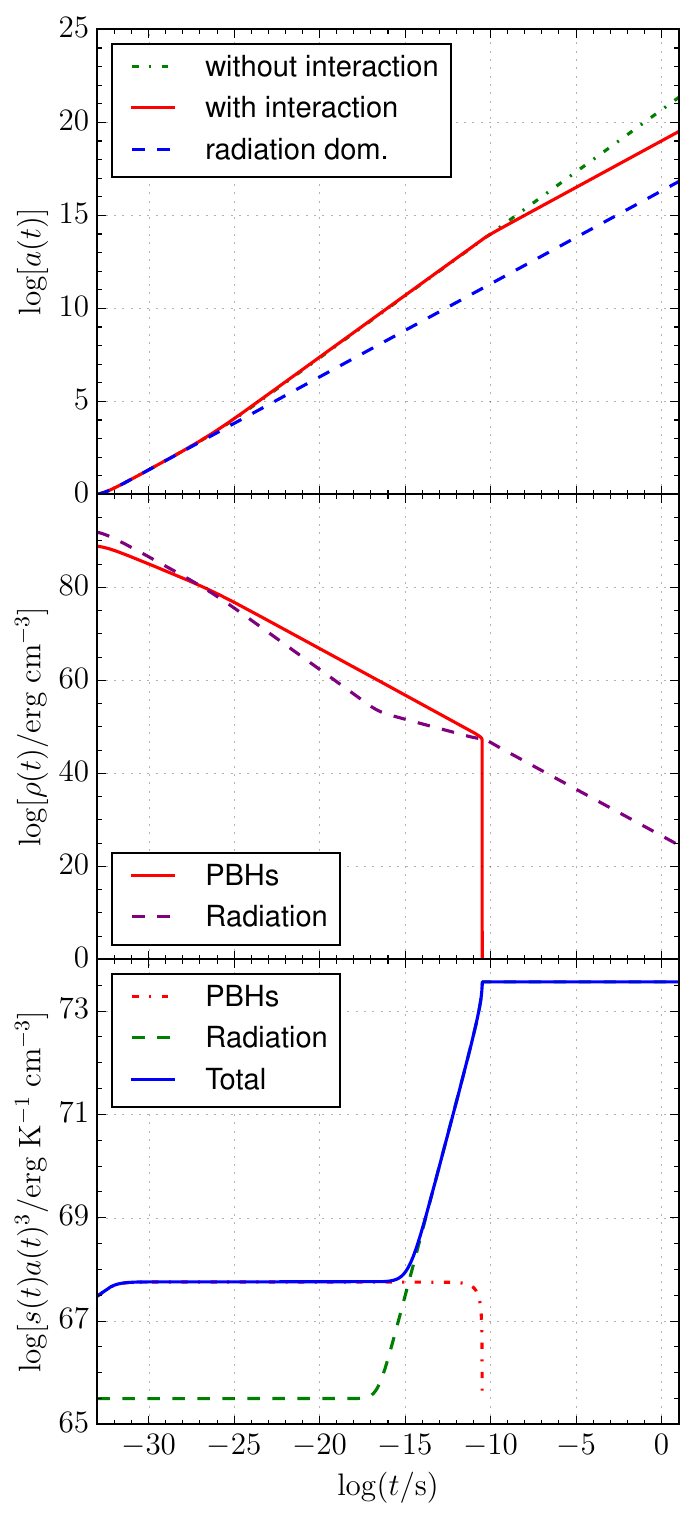}
	\caption{Cosmic evolution for a two-fluid model considering radiation and a monoenergetic population of PBHs with mass $M_{\rm PBH}=10^{5} \ \mathrm{g}$, and $\beta=10^{-3}$. Top panel shows the scale factor evolution for this system (with interaction), for a radiation-dominated universe (radiation), and for a universe in which black holes do not evaporate and behave exactly like dust (without interaction). Central panel shows the energy densities of both fluids, and bottom panel shows the entropy per unit of comoving volume of both fluids as well as the total one.}
\label{fig:DD}
\end{figure}

	Let us consider a monoenergetic PBH population of mass $M_{\rm ini} = 10^{5} ~ \mathrm{g} \sim M_{\rm H}(t_{\rm ini})$.
Figure \ref{fig:BH_mass} shows the evolution of one of these black holes.
	The only free parameter is the ratio of energy densities $\beta$.
In a first scenario, we do not take into account the rather speculative `relic particle' constraints. Hence, there is no upper limit on $\beta$ (for the chosen value of the mass) and we set $\beta = 10^{-3}$.
Figure \ref{fig:DD} shows the cosmic evolution for this scenario.
At the beginning, radiation dominates and the scale factor evolves as $a(t) \propto t^{1/2}$.
As the PBH component dilute slower than radiation, it starts to dominate at some later time, and during a period the universe is PBH-dominated (dust-dominated) and $a(t) \propto t^{2/3}$.
All black holes evaporate on a timescale of about $10^{-10} \ \mathrm{s}$ yielding their energy to radiation, and the universe becomes radiation-dominated again; however, the PBH population produced an increase in the scale factor of about two orders of magnitude.

The entropy in a comoving volume, $s(t)a(t)^3$, also increases during the whole evolution; this is driven by the accretion at early times and  by the evaporation at the end of the evolution.
The latter is the most significant process and produces an increase in entropy by a factor $\sim 10^6$. This scenario is an example of an \textit{a priori} plausible scenario (provided that the `relic particle' constraint does not apply) which presents significant modifications compared with the standard radiation-dominated evolution.

\begin{figure}
	\centering
	\includegraphics[width=0.4\textwidth,keepaspectratio]{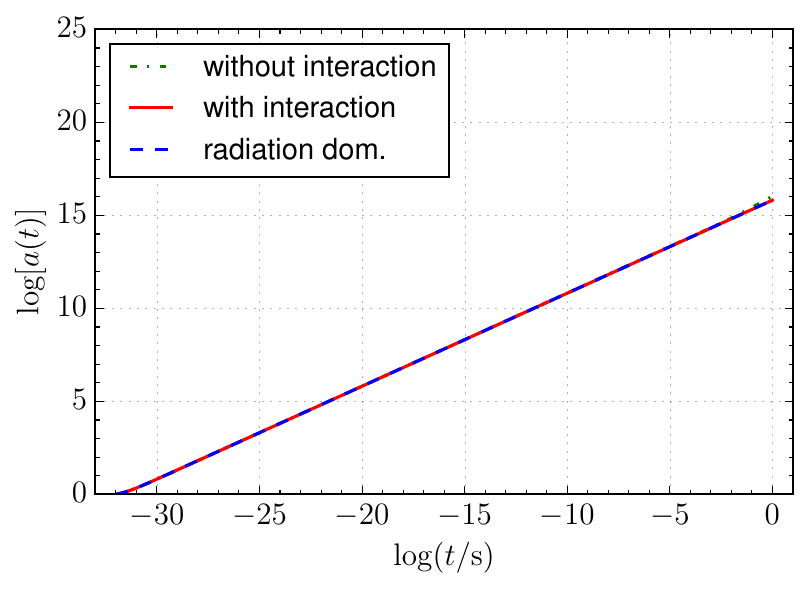}
	\caption{Scale factor evolution of a two-fluid model considering radiation and a population of PBHs with mass $M_{\rm PBH}=10^{5} \ \mathrm{g}$, and $\beta=10^{-15}$.}
\label{fig:DD_2}
\end{figure}

	If we accept the relic constraints, the highest possible value of $\beta$ for the adopted mass is $\sim 10^{-15}$. This is a similar case to the situation discussed, but with this new value of $\beta$ no modifications in the cosmic evolution are obtained, as it is shown in Fig. \ref{fig:DD_2}.

\subsection{Power-law IMF}

	Scenarios with a power-law IMF were studied by \citet{barrow1991}, who neglected the accretion and considered that all PBHs formed simultaneously.
We assume here that a black hole of mass $m$ cannot be formed until the horizon mass exceeds $m$, and hence the formation must be extended in time.
There are three additional free-parameters besides $\beta$: the minimum and maximum mass of the distribution, $M_{\rm min}$ y $M_{\rm max}$, and the spectral index $\gamma$.

	In order to investigate the general behaviour of power-law IMFs, we first ignore the constraints. We choose $M_{\rm min}=6.5 \times 10^2 m_{\rm P}$ ($m_{\rm P} \simeq 2.18 \times 10^{-5} \ \mathrm{g}$ is the Planck mass), which corresponds to black holes with lifetimes of the order of the initial time ($t_{\rm life} \sim t_{\rm ini}=10^{-33} ~ \mathrm{s}$), and $M_{\rm max}=10^6 ~ \rm M_\odot$, corresponding to supermassive black holes. We study IMFs with different values of $\gamma$ in the range $2-3$, and for different values of $\beta$. In addition, as we may have large enough PBHs for accretion to be important, we consider scenarios with and without accretion. We fit the scale factor with power-laws: $a \propto t^r$. 
	
\begin{figure}
	\centering
	\includegraphics[width=0.4\textwidth,keepaspectratio]{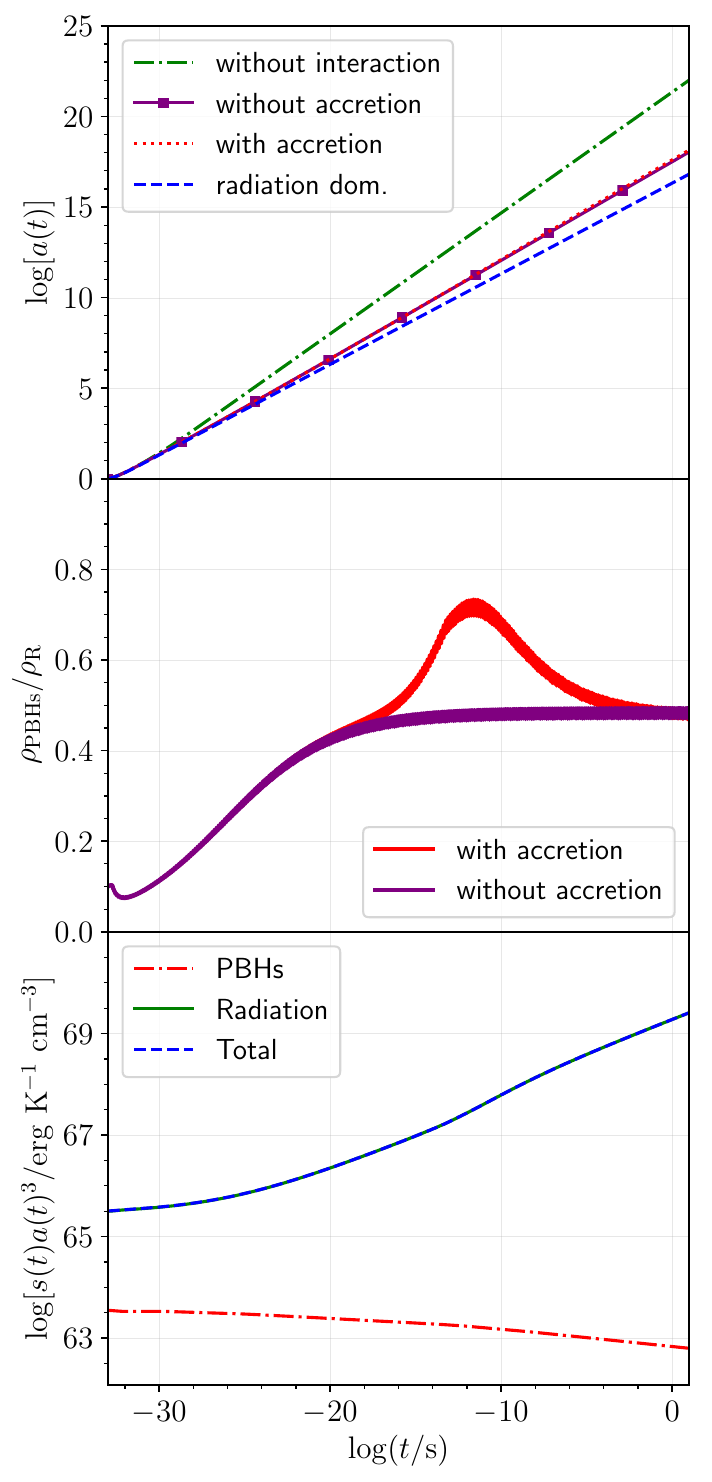}
	\caption{Cosmic evolution for a power-law IMF with spectral index $\gamma=2.1$, and with masses lying between $M_{\rm min}=650 m_{\rm P}$ and $M_{\rm max}=10^6 \ M_\odot$, for a value of $\beta=10^{-1}$. Top panel shows the scale factor evolution in four cases: with and without accretion, in a radiation dominated universe, and in a universe with radiation and PBHs but without any interaction between them. Central panel shows the ratio of the fluid energy densities with and without accretion. Bottom panel shows the entropy per unit of comoving volume of both fluids as well as the total one.}	
\label{fig:pl_1}
\end{figure}

In agreement with \citet{barrow1992}, we find scenarios with intermediate evolution between those that are radiation and dust-dominated ($1/2 < r < 2/3$).
Significant modifications in the cosmic evolution only occur for $\gamma$ near $2$ (harder spectra) and for higher values of $\beta$.
Figure \ref{fig:pl_1} shows the results for the case $\gamma=2.1$, $\beta=10^{-1}$. The scale factor evolves as a power-law with $r=0.544$ independently of whether there is accretion or not. Furthermore, despite accretion temporarily modifies the ratio of energy densities, the dilution of the holes rapidly becomes dominant and the two scenarios converge to the same stationary value of $\beta \sim 0.47$; accretion seems to be negligible even in the most favourable scenario. Finally, the entropy in a comoving volume increases during the whole evolution although less than in one of the monoenergetic cases discussed above, even though here $\beta$ is two orders of magnitude higher.

\begin{figure}
	\centering
	\includegraphics[width=0.4\textwidth,keepaspectratio]{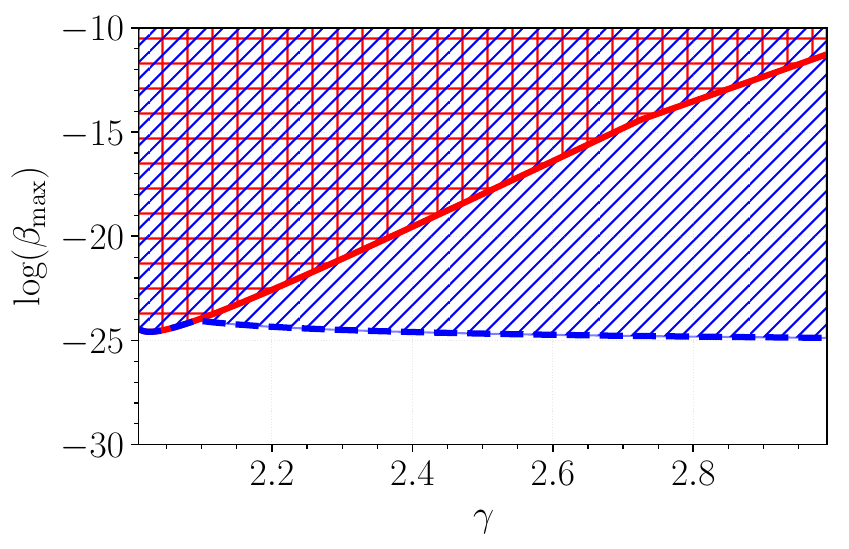}
	\caption{ Combined constraints on the initial fraction of the energy density of the universe in PBHs for power-law IMFs, as a function of the spectral index $\gamma$. The minimum mass is $M_\mathrm{min}=650 m_{\rm P}$. The red line is plotted discarding the relic constraints whereas the blue dashed-line considers them. The hatched areas are the forbidden regions of the parameter $\beta$. In the Appendix \ref{ap:constraints_extended} we show a more general case.}
\label{fig:constraints_ef}
\end{figure}

	A plausible power-law scenario must satisfy the observational constraints. Figure \ref{fig:constraints_ef} shows how these constraints apply for this type of distribution as a function of $\gamma$. We can see that the harder the spectrum is, the lower the influence of the relic constraints results. The most favourable scenarios are those with hard spectra; for them $\beta \sim 10^{-24}$ both neglecting and including the relic constraints. For these cases, we obtain that PBHs do not cause any significant effect and the scale factor evolves as in a radiation-dominated universe.
	
\section{Conclusions and Final Remarks}\label{Sec:conclusions}

	We have studied the effects that different PBH populations produce on the evolution of the early universe. We considered a two-fluid cosmological model with radiation and a PBH gas. According to the different possible formation mechanisms, the PBH IMF may be extended or narrow. We have investigated representative cases for these two possible scenarios, namely power-law and monoenergetic IMFs.
To characterise the IMFs, we have taken into account the different constraints that exist for PBH abundances.
In particular, we have distinguished the `relic constraints' from the others, since they are too conjectural. 

	Monoenergetic populations of small PBHs produce significant modifications in the cosmic evolution (changes in the scale factor and generation of entropy) provided that the PBH energy density is high enough. The latter condition requires to discard the `relic constraints'. If instead we take them as valid, no effects are produced.
	
	The behaviour of power-law scenarios is different. We discussed that the constraint treatment depends on the IMF form. For the analysed cases, we found cosmic evolution modifications only if we omit the constraints. In addition, we investigated the importance of accretion in these scenarios finding that it plays no significant role.
Situations where all constraints are satisfied do not present any relevant effect.

	We conclude that the presence of some particular PBH populations in the early universe may affect the cosmic evolution. In particular, populations of small PBHs with narrow IMFs are likely to produce these effects. These populations are possible if the relic constraints are not valid. In addition, we found that accretion on to PBHs is not a relevant energy-exchange mechanism in the early universe, even for the most favourable IMFs.

\appendix
\section{PBH constraints on extended mass functions}\label{ap:constraints_extended}
	Let us consider a PBH population with mass function $N(m; \alpha )$, where $\alpha =  \{ \alpha_i \}$ are the parameters that characterise the function, and let us define
\begin{equation}
	\phi (m;\alpha) := \rho_\mathrm{tot}^{-1}N(m;\alpha) mc^2,
\end{equation} 
where $\rho_\mathrm{tot}$ is the total energy density of the universe and $c$ is the speed of light in vacuum. If $\mathcal{O}[\phi (m;\alpha)]$ denotes an observable depending on the PBH mass function, we can expand it as
\begin{equation}
\begin{aligned}
	\mathcal{O}[\phi (m;\alpha)] = \mathcal{O}_0 + \int dm \phi (m;\alpha) K_1(m) + \\
	 \int dm_1 dm_2 \phi (m_1;\alpha) \phi (m_2;\alpha) K_2 (m_1, m_2) + ...,
\end{aligned}
\label{eq:A}
\end{equation}
where $\mathcal{O}_0$ is the background contribution and the functions $K_j(m)$ depend on the details of the underlying physics and the nature of the observation. As we considered that PBHs do not interact among themselves, only the first two terms in the right-hand side of Eq. (\ref{eq:A}) need to be considered.

\begin{figure}
	\centering
	\includegraphics[width=0.48\textwidth,keepaspectratio]{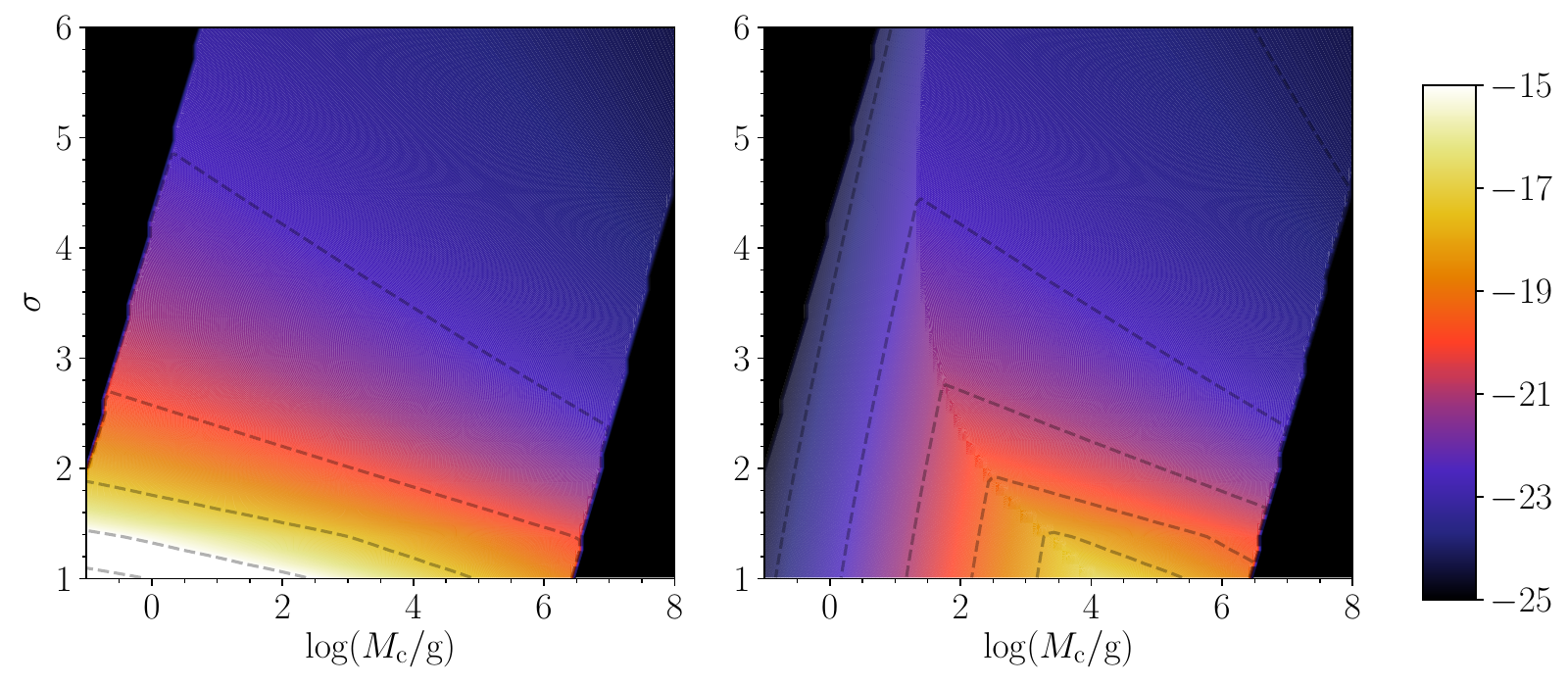}
	\caption{Combined effect of the constraints on the initial fraction of the energy density of the universe in PBHs for power-law IMFs, as a function of the parameters $M_\mathrm{c}=M_\mathrm{min} e^{1/(\gamma-2)}$ and $\sigma \equiv 1/(\gamma - 2)$. The colour map shows $\log (\beta_\mathrm{max})$. The dotted-lines are contour lines of $\beta_\mathrm{max}$ (i.e., $\beta_\mathrm{max}=\mathrm{const.}$) and the regions in black are forbidden combinations of the parameters. The left panel is plotted discarding the relic constraints whereas in the right panel these are considered.}
\label{fig:constraints_ef_complete}
\end{figure}

	If a measurement imposes an upper bound on the observable,
\begin{equation}
	\mathcal{O}[\phi (m;\alpha)] \leq \mathcal{O}_\mathrm{exp},
\end{equation}
for a monoenergetic mass function with $m=M_*$,
\begin{equation}
	\phi_\mathrm{mon} (M_*) \equiv \rho_\mathrm{tot}^{-1} A \delta (m-M_*) m c^2,
\end{equation}
we have
\begin{equation}
	\rho_\mathrm{tot}^{-1} A M_* c^2  \leq \frac{\mathcal{O}_\mathrm{exp} - \mathcal{O}_0}{K_1(M_*)} \equiv \beta (M_*),
\label{eq:A_mono}
\end{equation}
where $\beta (M_*)$ is the upper bound for monoenergetic distributions (see Fig. \ref{fig:constraints}).
Combining Eqs. (\ref{eq:A}) and (\ref{eq:A_mono}) we obtain
\begin{equation}
	\int dm \frac{\phi (m;\alpha)}{\beta (m)} \leq 1,
\label{eq:constraints_arbitrary}
\end{equation}
for an arbitrary mass function.
If we know $\beta(m)$ and assume the form of the function $\phi (m;\alpha)$, we can integrate Eq. (\ref{eq:constraints_arbitrary}) over the mass range $(m_1, m_2)$ for which the constraint applies. For given values of the parameters $\alpha$, this imposes limits on $\beta \equiv \rho_\mathrm{PBH}/\rho_\mathrm{tot}$.
In particular, for a power-law mass function we have
\begin{equation}
	\phi (m; M_\mathrm{min}, \gamma) = A m^{1-\gamma} c^2, ~ \mathrm{with} ~ m \geq M_\mathrm{min}.
\end{equation}
If $\gamma>2$, a minimum mass is strictly necessary for the function not to diverge. Instead, the role of the maximum mass is not important in this case. Then, for each set of parameters $\{\gamma, M_\mathrm{min}\}$ the combined effect of Eq. (\ref{eq:constraints_arbitrary}) applied to the different constraints impose limits on $A$, and hence on $\beta$. It is standard to plot the constraints as a function of the parameters $M_\mathrm{c} \equiv M_\mathrm{min} e^{1/(\gamma-2)}$ and $\sigma \equiv 1/(\gamma - 2)$ instead of the original ones (see Fig. \ref{fig:constraints_ef_complete}).

The procedure discussed in this Appendix is an adaptation of the one presented by \citet{carr2017}. 

\section*{Acknowledgments}

This work was supported by the Argentine Agency CONICET (PIP 2014-00338) and the Spanish Ministerio de Econom\'{i}a y Competitividad (MINECO/FEDER, UE) under grants AYA2013-47447-C3-1-P and AYA2016-76012-C3-1-P.

\bibliographystyle{mnras}
\bibliography{refs}   

\begin{thebibliography}{}
\makeatletter
\relax
\def\mn@urlcharsother{\let\do\@makeother \do\$\do\&\do\#\do\^\do\_\do\%\do\~}
\def\mn@doi{\begingroup\mn@urlcharsother \@ifnextchar [ {\mn@doi@}
  {\mn@doi@[]}}
\def\mn@doi@[#1]#2{\def\@tempa{#1}\ifx\@tempa\@empty \href
  {http://dx.doi.org/#2} {doi:#2}\else \href {http://dx.doi.org/#2} {#1}\fi
  \endgroup}
\def\mn@eprint#1#2{\mn@eprint@#1:#2::\@nil}
\def\mn@eprint@arXiv#1{\href {http://arxiv.org/abs/#1} {{\tt arXiv:#1}}}
\def\mn@eprint@dblp#1{\href {http://dblp.uni-trier.de/rec/bibtex/#1.xml}
  {dblp:#1}}
\def\mn@eprint@#1:#2:#3:#4\@nil{\def\@tempa {#1}\def\@tempb {#2}\def\@tempc
  {#3}\ifx \@tempc \@empty \let \@tempc \@tempb \let \@tempb \@tempa \fi \ifx
  \@tempb \@empty \def\@tempb {arXiv}\fi \@ifundefined
  {mn@eprint@\@tempb}{\@tempb:\@tempc}{\expandafter \expandafter \csname
  mn@eprint@\@tempb\endcsname \expandafter{\@tempc}}}

\bibitem[\protect\citeauthoryear{Abbott et~al.}{Abbott
  et~al.}{2016a}]{abbott2016a}
Abbott B.~P.,  et~al., 2016a, \mn@doi [Phys. Rev. Lett.]
  {10.1103/PhysRevLett.116.061102}, 116, 061102

\bibitem[\protect\citeauthoryear{Abbott et~al.}{Abbott
  et~al.}{2016b}]{abbott2016b}
Abbott B.~P.,  et~al., 2016b, \mn@doi [Phys. Rev. Lett.]
  {10.1103/PhysRevLett.116.241103}, 116, 241103

\bibitem[\protect\citeauthoryear{Abbott et~al.}{Abbott
  et~al.}{2017}]{abbott2017}
Abbott B.~P.,  et~al., 2017, \mn@doi [Phys. Rev. Lett.]
  {10.1103/PhysRevLett.118.221101}, 118, 221101

\bibitem[\protect\citeauthoryear{{Akerib} et~al.,}{{Akerib}
  et~al.}{2016}]{akerib2016}
{Akerib} D.~S.,  et~al., 2016, \mn@doi [Physical Review Letters]
  {10.1103/PhysRevLett.116.161301}, \href
  {http://adsabs.harvard.edu/abs/2016PhRvL.116p1301A} {116, 161301}

\bibitem[\protect\citeauthoryear{{Alexander} \& {M{\'e}sz{\'a}ros}}{{Alexander}
  \& {M{\'e}sz{\'a}ros}}{2007}]{alexander2007}
{Alexander} S.,  {M{\'e}sz{\'a}ros} P.,  2007, [arXiv:2007hep-th/0703070A,
  \href {http://adsabs.harvard.edu/abs/2007hep.th....3070A} {}

\bibitem[\protect\citeauthoryear{{Barrow}}{{Barrow}}{1980}]{barrow1980}
{Barrow} J.~D.,  1980, \mn@doi [\mnras] {10.1093/mnras/192.3.427}, \href
  {http://adsabs.harvard.edu/abs/1980MNRAS.192..427B} {192, 427}

\bibitem[\protect\citeauthoryear{{Barrow}, {Copeland}, {Kolb}  \&
  {Liddle}}{{Barrow} et~al.}{1991a}]{barrow1991b}
{Barrow} J.~D.,  {Copeland} E.~J.,  {Kolb} E.~W.,   {Liddle} A.~R.,  1991a,
  \mn@doi [\prd] {10.1103/PhysRevD.43.984}, \href
  {http://adsabs.harvard.edu/abs/1991PhRvD..43..984B} {43, 984}

\bibitem[\protect\citeauthoryear{{Barrow}, {Copeland}  \& {Liddle}}{{Barrow}
  et~al.}{1991b}]{barrow1991}
{Barrow} J.~D.,  {Copeland} E.~J.,   {Liddle} A.~R.,  1991b, \mn@doi [\mnras]
  {10.1093/mnras/253.4.675}, \href
  {http://adsabs.harvard.edu/abs/1991MNRAS.253..675B} {253, 675}

\bibitem[\protect\citeauthoryear{{Barrow}, {Copeland}  \& {Liddle}}{{Barrow}
  et~al.}{1992}]{barrow1992}
{Barrow} J.~D.,  {Copeland} E.~J.,   {Liddle} A.~R.,  1992, \mn@doi [\prd]
  {10.1103/PhysRevD.46.645}, \href
  {http://adsabs.harvard.edu/abs/1992PhRvD..46..645B} {46, 645}

\bibitem[\protect\citeauthoryear{{Berezin}, {Kuzmin}  \& {Tkachev}}{{Berezin}
  et~al.}{1983}]{berezin1983}
{Berezin} V.~A.,  {Kuzmin} V.~A.,   {Tkachev} I.~I.,  1983, \mn@doi [Physics
  Letters B] {10.1016/0370-2693(83)90630-5}, \href
  {http://adsabs.harvard.edu/abs/1983PhLB..120...91B} {120, 91}

\bibitem[\protect\citeauthoryear{{Bird}, {Cholis}, {Mu{\~n}oz},
  {Ali-Ha{\"i}moud}, {Kamionkowski}, {Kovetz}, {Raccanelli}  \& {Riess}}{{Bird}
  et~al.}{2016}]{bird2016}
{Bird} S.,  {Cholis} I.,  {Mu{\~n}oz} J.~B.,  {Ali-Ha{\"i}moud} Y.,
  {Kamionkowski} M.,  {Kovetz} E.~D.,  {Raccanelli} A.,   {Riess} A.~G.,  2016,
  \mn@doi [Physical Review Letters] {10.1103/PhysRevLett.116.201301}, \href
  {http://adsabs.harvard.edu/abs/2016PhRvL.116t1301B} {116, 201301}

\bibitem[\protect\citeauthoryear{{Borunda} \& {Masip}}{{Borunda} \&
  {Masip}}{2010}]{borunda2010}
{Borunda} M.,  {Masip} M.,  2010, \mn@doi [\jcap]
  {10.1088/1475-7516/2010/01/027}, \href
  {http://adsabs.harvard.edu/abs/2010JCAP...01..027B} {1, 027}

\bibitem[\protect\citeauthoryear{{Brevik} \& {Halnes}}{{Brevik} \&
  {Halnes}}{2003}]{brevik2003}
{Brevik} I.,  {Halnes} G.,  2003, \mn@doi [\prd] {10.1103/PhysRevD.67.023508},
  \href {http://adsabs.harvard.edu/abs/2003PhRvD..67b3508B} {67, 023508}

\bibitem[\protect\citeauthoryear{{Bugaev} \& {Konishchev}}{{Bugaev} \&
  {Konishchev}}{2002}]{bugaev2002}
{Bugaev} E.~V.,  {Konishchev} K.~V.,  2002, \mn@doi [\prd]
  {10.1103/PhysRevD.66.084004}, \href
  {http://adsabs.harvard.edu/abs/2002PhRvD..66h4004B} {66, 084004}

\bibitem[\protect\citeauthoryear{{Bugaev}, {Elbakidze}  \&
  {Konishche}}{{Bugaev} et~al.}{2003}]{bugaev2003}
{Bugaev} E.~V.,  {Elbakidze} M.~G.,   {Konishche} K.~V.,  2003, \mn@doi
  [Physics of Atomic Nuclei] {10.1134/1.1563709}, \href
  {http://adsabs.harvard.edu/abs/2003PAN....66..476B} {66, 476}

\bibitem[\protect\citeauthoryear{{Caldwell}, {Chamblin}  \&
  {Gibbons}}{{Caldwell} et~al.}{1996}]{caldwell1996}
{Caldwell} R.~R.,  {Chamblin} H.~A.,   {Gibbons} G.~W.,  1996, \mn@doi [\prd]
  {10.1103/PhysRevD.53.7103}, \href
  {http://adsabs.harvard.edu/abs/1996PhRvD..53.7103C} {53, 7103}

\bibitem[\protect\citeauthoryear{{Capela}, {Pshirkov}  \& {Tinyakov}}{{Capela}
  et~al.}{2013}]{capela2013}
{Capela} F.,  {Pshirkov} M.,   {Tinyakov} P.,  2013, \mn@doi [\prd]
  {10.1103/PhysRevD.87.123524}, \href
  {http://adsabs.harvard.edu/abs/2013PhRvD..87l3524C} {87, 123524}

\bibitem[\protect\citeauthoryear{{Carr}}{{Carr}}{1975}]{carr1975}
{Carr} B.~J.,  1975, \mn@doi [\apj] {10.1086/153853}, \href
  {http://adsabs.harvard.edu/abs/1975ApJ...201....1C} {201, 1}

\bibitem[\protect\citeauthoryear{{Carr} \& {Hawking}}{{Carr} \&
  {Hawking}}{1974}]{carr1974}
{Carr} B.~J.,  {Hawking} S.~W.,  1974, \mn@doi [\mnras]
  {10.1093/mnras/168.2.399}, \href
  {http://adsabs.harvard.edu/abs/1974MNRAS.168..399C} {168, 399}

\bibitem[\protect\citeauthoryear{{Carr}, {Kohri}, {Sendouda}  \&
  {Yokoyama}}{{Carr} et~al.}{2010}]{carr2010}
{Carr} B.~J.,  {Kohri} K.,  {Sendouda} Y.,   {Yokoyama} J.,  2010, \mn@doi
  [\prd] {10.1103/PhysRevD.81.104019}, \href
  {http://adsabs.harvard.edu/abs/2010PhRvD..81j4019C} {81, 104019}

\bibitem[\protect\citeauthoryear{{Carr}, {Kohri}, {Sendouda}  \&
  {Yokoyama}}{{Carr} et~al.}{2016a}]{carr2016b}
{Carr} B.~J.,  {Kohri} K.,  {Sendouda} Y.,   {Yokoyama} J.,  2016a, \mn@doi
  [\prd] {10.1103/PhysRevD.94.044029}, \href
  {http://adsabs.harvard.edu/abs/2016PhRvD..94d4029C} {94, 044029}

\bibitem[\protect\citeauthoryear{{Carr}, {K{\"u}hnel}  \& {Sandstad}}{{Carr}
  et~al.}{2016b}]{carr2016a}
{Carr} B.,  {K{\"u}hnel} F.,   {Sandstad} M.,  2016b, \mn@doi [\prd]
  {10.1103/PhysRevD.94.083504}, \href
  {http://adsabs.harvard.edu/abs/2016PhRvD..94h3504C} {94, 083504}

\bibitem[\protect\citeauthoryear{{Carr}, {Raidal}, {Tenkanen}, {Vaskonen}  \&
  {Veerm{\"a}e}}{{Carr} et~al.}{2017}]{carr2017}
{Carr} B.,  {Raidal} M.,  {Tenkanen} T.,  {Vaskonen} V.,   {Veerm{\"a}e} H.,
  2017, \prd, \href {http://adsabs.harvard.edu/abs/2017arXiv170505567C} {96}

\bibitem[\protect\citeauthoryear{{Chapline}}{{Chapline}}{1975}]{chapline1975}
{Chapline} G.~F.,  1975, \mn@doi [\nat] {10.1038/253251a0}, \href
  {http://adsabs.harvard.edu/abs/1975Natur.253..251C} {253, 251}

\bibitem[\protect\citeauthoryear{{Corral-Santana}, {Casares},
  {Mu{\~n}oz-Darias}, {Bauer}, {Mart{\'{\i}}nez-Pais}  \&
  {Russell}}{{Corral-Santana} et~al.}{2016}]{corral2016}
{Corral-Santana} J.~M.,  {Casares} J.,  {Mu{\~n}oz-Darias} T.,  {Bauer} F.~E.,
  {Mart{\'{\i}}nez-Pais} I.~G.,   {Russell} D.~M.,  2016, \mn@doi [\aap]
  {10.1051/0004-6361/201527130}, \href
  {http://adsabs.harvard.edu/abs/2016A%26A...587A..61C} {587, A61}

\bibitem[\protect\citeauthoryear{Crawford \& Schramm}{Crawford \&
  Schramm}{1982}]{crawford1982}
Crawford M.,  Schramm D.~N.,  1982, \mn@doi [Nature] {10.1038/298538a0}, 298,
  538

\bibitem[\protect\citeauthoryear{{Dolgov}, {Naselsky}  \& {Novikov}}{{Dolgov}
  et~al.}{2000}]{dolgov2000}
{Dolgov} A.~D.,  {Naselsky} P.~D.,   {Novikov} I.~D.,  2000,
  [2000astro.ph..9407D], \href
  {http://adsabs.harvard.edu/abs/2000astro.ph..9407D} {}

\bibitem[\protect\citeauthoryear{{Ferrarese} \& {Ford}}{{Ferrarese} \&
  {Ford}}{2005}]{ferrarese2005}
{Ferrarese} L.,  {Ford} H.,  2005, \mn@doi [{Space Science Reviews}]
  {10.1007/s11214-005-3947-6}, \href
  {http://adsabs.harvard.edu/abs/2005SSRv..116..523F} {116, 523}

\bibitem[\protect\citeauthoryear{{Garc{\'{\i}}a-Bellido}}{{Garc{\'{\i}}a-Bellido}}{2017}]{garcia2017}
{Garc{\'{\i}}a-Bellido} J.,  2017, in Journal of Physics Conference Series. p.
  012032 (\mn@eprint {arXiv} {1702.08275}),
  \mn@doi{10.1088/1742-6596/840/1/012032}

\bibitem[\protect\citeauthoryear{{Gibilisco}}{{Gibilisco}}{1998}]{gibilisco1998}
{Gibilisco} M.,  1998, in {Coccia} E.,  {Veneziano} G.,   {Pizzella} G.,  eds,
  Second Edoardo Amaldi Conference on Gravitational Wave Experiments. p.~314
  (\mn@eprint {} {astro-ph/9709297})

\bibitem[\protect\citeauthoryear{Green}{Green}{2015}]{green2014}
Green A.~M.,  2015, \mn@doi [Fundam. Theor. Phys.]
  {10.1007/978-3-319-10852-0_5}, 178, 129

\bibitem[\protect\citeauthoryear{{Green} \& {Liddle}}{{Green} \&
  {Liddle}}{1999}]{green1999}
{Green} A.~M.,  {Liddle} A.~R.,  1999, \mn@doi [\prd]
  {10.1103/PhysRevD.60.063509}, \href
  {http://adsabs.harvard.edu/abs/1999PhRvD..60f3509G} {60, 063509}

\bibitem[\protect\citeauthoryear{{Hawking}}{{Hawking}}{1971}]{hawking1971}
{Hawking} S.,  1971, \mn@doi [\mnras] {10.1093/mnras/152.1.75}, \href
  {http://adsabs.harvard.edu/abs/1971MNRAS.152...75H} {152, 75}

\bibitem[\protect\citeauthoryear{{Hawking}}{{Hawking}}{1974}]{hawking1974}
{Hawking} S.~W.,  1974, \mn@doi [\nat] {10.1038/248030a0}, \href
  {http://adsabs.harvard.edu/abs/1974Natur.248...30H} {248, 30}

\bibitem[\protect\citeauthoryear{{Hawking}}{{Hawking}}{1975}]{hawking1975}
{Hawking} S.~W.,  1975, \mn@doi [Communications in Mathematical Physics]
  {10.1007/BF02345020}, \href
  {http://adsabs.harvard.edu/abs/1975CMaPh..43..199H} {43, 199}

\bibitem[\protect\citeauthoryear{Hawking}{Hawking}{1987}]{hawking1987}
Hawking S.~W.,  1987, \mn@doi [Phys. Lett.] {10.1016/0370-2693(89)90206-2},
  B231, 237

\bibitem[\protect\citeauthoryear{Hawking, Moss  \& Stewart}{Hawking
  et~al.}{1982}]{hawking1982}
Hawking S.~W.,  Moss I.~G.,   Stewart J.~M.,  1982, \mn@doi [Phys. Rev.]
  {10.1103/PhysRevD.26.2681}, D26, 2681

\bibitem[\protect\citeauthoryear{{Hook}}{{Hook}}{2014}]{hook2014}
{Hook} A.,  2014, \mn@doi [\prd] {10.1103/PhysRevD.90.083535}, \href
  {http://adsabs.harvard.edu/abs/2014PhRvD..90h3535H} {90, 083535}

\bibitem[\protect\citeauthoryear{{Jedamzik}}{{Jedamzik}}{1997}]{jedamzik1997}
{Jedamzik} K.,  1997, \mn@doi [\prd] {10.1103/PhysRevD.55.R5871}, \href
  {http://adsabs.harvard.edu/abs/1997PhRvD..55.5871J} {55, R5871}

\bibitem[\protect\citeauthoryear{{Jedamzik} \& {Niemeyer}}{{Jedamzik} \&
  {Niemeyer}}{1999}]{jedamzik1999}
{Jedamzik} K.,  {Niemeyer} J.~C.,  1999, \mn@doi [\prd]
  {10.1103/PhysRevD.59.124014}, \href
  {http://adsabs.harvard.edu/abs/1999PhRvD..59l4014J} {59, 124014}

\bibitem[\protect\citeauthoryear{{Josan} \& {Green}}{{Josan} \&
  {Green}}{2010}]{josan2010}
{Josan} A.~S.,  {Green} A.~M.,  2010, \mn@doi [\prd]
  {10.1103/PhysRevD.82.047303}, \href
  {http://adsabs.harvard.edu/abs/2010PhRvD..82d7303J} {82, 047303}

\bibitem[\protect\citeauthoryear{{Khlopov}}{{Khlopov}}{2010}]{khlopov2010}
{Khlopov} M.~Y.,  2010, \mn@doi [Research in Astronomy and Astrophysics]
  {10.1088/1674-4527/10/6/001}, \href
  {http://adsabs.harvard.edu/abs/2010RAA....10..495K} {10, 495}

\bibitem[\protect\citeauthoryear{{Khlopov} \& {Polnarev}}{{Khlopov} \&
  {Polnarev}}{1980}]{khlopov1980}
{Khlopov} M.~Y.,  {Polnarev} A.~G.,  1980, \mn@doi [Physics Letters B]
  {10.1016/0370-2693(80)90624-3}, \href
  {http://adsabs.harvard.edu/abs/1980PhLB...97..383K} {97, 383}

\bibitem[\protect\citeauthoryear{{Kohri} \& {Yokoyama}}{{Kohri} \&
  {Yokoyama}}{2000}]{kohri2000}
{Kohri} K.,  {Yokoyama} J.,  2000, \mn@doi [\prd] {10.1103/PhysRevD.61.023501},
  \href {http://adsabs.harvard.edu/abs/2000PhRvD..61b3501K} {61, 023501}

\bibitem[\protect\citeauthoryear{{K{\"u}hnel} \& {Freese}}{{K{\"u}hnel} \&
  {Freese}}{2017}]{kuhnel2017}
{K{\"u}hnel} F.,  {Freese} K.,  2017, \mn@doi [\prd]
  {10.1103/PhysRevD.95.083508}, \href
  {http://adsabs.harvard.edu/abs/2017PhRvD..95h3508K} {95, 083508}

\bibitem[\protect\citeauthoryear{{Lemoine}}{{Lemoine}}{2000}]{lemoine2000}
{Lemoine} M.,  2000, \mn@doi [Physics Letters B]
  {10.1016/S0370-2693(00)00469-X}, \href
  {http://adsabs.harvard.edu/abs/2000PhLB..481..333L} {481, 333}

\bibitem[\protect\citeauthoryear{{Lindley}}{{Lindley}}{1981}]{lindley1981}
{Lindley} D.,  1981, \mn@doi [\mnras] {10.1093/mnras/196.2.317}, \href
  {http://adsabs.harvard.edu/abs/1981MNRAS.196..317L} {196, 317}

\bibitem[\protect\citeauthoryear{{MacGibbon}}{{MacGibbon}}{1987}]{macgibbon1987}
{MacGibbon} J.~H.,  1987, \mn@doi [\nat] {10.1038/329308a0}, \href
  {http://adsabs.harvard.edu/abs/1987Natur.329..308M} {329, 308}

\bibitem[\protect\citeauthoryear{{MacGibbon} \& {Carr}}{{MacGibbon} \&
  {Carr}}{1991}]{macgibbon1991}
{MacGibbon} J.~H.,  {Carr} B.~J.,  1991, \mn@doi [\apj] {10.1086/169909}, \href
  {http://adsabs.harvard.edu/abs/1991ApJ...371..447M} {371, 447}

\bibitem[\protect\citeauthoryear{{Neugebauer}}{{Neugebauer}}{2003}]{neugebauer2003}
{Neugebauer} G.,  2003, {The collapse to a black hole}.
{Falcke}, H. and {Hehl}, F.~W., pp 72--94

\bibitem[\protect\citeauthoryear{{Niemeyer} \& {Jedamzik}}{{Niemeyer} \&
  {Jedamzik}}{1998}]{niemeyer1998}
{Niemeyer} J.~C.,  {Jedamzik} K.,  1998, \mn@doi [Physical Review Letters]
  {10.1103/PhysRevLett.80.5481}, \href
  {http://adsabs.harvard.edu/abs/1998PhRvL..80.5481N} {80, 5481}

\bibitem[\protect\citeauthoryear{{Peiris} \& {Easther}}{{Peiris} \&
  {Easther}}{2008}]{peiris2008}
{Peiris} H.~V.,  {Easther} R.,  2008, \mn@doi [\jcap]
  {10.1088/1475-7516/2008/07/024}, \href
  {http://adsabs.harvard.edu/abs/2008JCAP...07..024P} {7, 024}

\bibitem[\protect\citeauthoryear{{Polnarev} \& {Khlopov}}{{Polnarev} \&
  {Khlopov}}{1985}]{polnarev1985}
{Polnarev} A.~G.,  {Khlopov} M.~Y.,  1985, \mn@doi [Soviet Physics Uspekhi]
  {10.1070/PU1985v028n03ABEH003858}, \href
  {http://adsabs.harvard.edu/abs/1985SvPhU..28..213P} {28, 213}

\bibitem[\protect\citeauthoryear{Polnarev \& Zembowicz}{Polnarev \&
  Zembowicz}{1991}]{polnarev1991}
Polnarev A.,  Zembowicz R.,  1991, \mn@doi [Phys. Rev.]
  {10.1103/PhysRevD.43.1106}, D43, 1106

\bibitem[\protect\citeauthoryear{{Sasaki}, {Suyama}, {Tanaka}  \&
  {Yokoyama}}{{Sasaki} et~al.}{2016}]{sasaki2016}
{Sasaki} M.,  {Suyama} T.,  {Tanaka} T.,   {Yokoyama} S.,  2016, \mn@doi
  [Physical Review Letters] {10.1103/PhysRevLett.117.061101}, \href
  {http://adsabs.harvard.edu/abs/2016PhRvL.117f1101S} {117, 061101}

\bibitem[\protect\citeauthoryear{{Vainer}, {Dryzhakova}  \&
  {Naselskii}}{{Vainer} et~al.}{1978}]{vainer1978}
{Vainer} B.~V.,  {Dryzhakova} O.~V.,   {Naselskii} P.~D.,  1978, Soviet
  Astronomy Letters, \href {http://adsabs.harvard.edu/abs/1978SvAL....4..185V}
  {4, 185}

\bibitem[\protect\citeauthoryear{Weinberg}{Weinberg}{1972}]{weinberg1972}
Weinberg S.,  1972, Gravitation and cosmology: principles and applications of
  the general theory of relativity.
Wiley, \url {https://books.google.com.ar/books?id=XLbvAAAAMAAJ}

\bibitem[\protect\citeauthoryear{{Wright}}{{Wright}}{1996}]{wright1996}
{Wright} E.~L.,  1996, \mn@doi [\apj] {10.1086/176910}, \href
  {http://adsabs.harvard.edu/abs/1996ApJ...459..487W} {459, 487}

\bibitem[\protect\citeauthoryear{{Zel'dovich} \& {Novikov}}{{Zel'dovich} \&
  {Novikov}}{1966}]{zeldovich1966}
{Zel'dovich} Y.~B.,  {Novikov} I.~D.,  1966, Astronomicheskii Zhurnal, \href
  {http://adsabs.harvard.edu/abs/1966AZh....43..758Z} {43, 758}

\bibitem[\protect\citeauthoryear{{Zeldovich} \& {Starobinskii}}{{Zeldovich} \&
  {Starobinskii}}{1976}]{zeldovich1976}
{Zeldovich} I.~B.,  {Starobinskii} A.~A.,  1976, ZhETF Pisma Redaktsiiu, \href
  {http://adsabs.harvard.edu/abs/1976ZhPmR..24..616Z} {24, 616}

\bibitem[\protect\citeauthoryear{{Zeldovich}, {Starobinskii}, {Khlopov}  \&
  {Chechetkin}}{{Zeldovich} et~al.}{1977}]{zeldovich1977}
{Zeldovich} I.~B.,  {Starobinskii} A.~A.,  {Khlopov} M.~I.,   {Chechetkin}
  V.~M.,  1977, Pisma v Astronomicheskii Zhurnal, \href
  {http://adsabs.harvard.edu/abs/1977PAZh....3..208Z} {3, 208}

\bibitem[\protect\citeauthoryear{{Zimdahl} \& {Pav{\'o}n}}{{Zimdahl} \&
  {Pav{\'o}n}}{1998}]{zimdahl1998}
{Zimdahl} W.,  {Pav{\'o}n} D.,  1998, \mn@doi [\prd]
  {10.1103/PhysRevD.58.103506}, \href
  {http://adsabs.harvard.edu/abs/1998PhRvD..58j3506Z} {58, 103506}

\makeatother
\end{thebibliography}




\label{lastpage}

\end{document}